\documentstyle[]{mn}

\begin{document}

\title{Toward an  Adequate  Method  to  Isolate  Spectroscopic
              Families of Diffuse Interstellar Bands}
 
\author[B.Wszo\l ek \& W. God\l owski]
{B.Wszo\l ek$^1$,$^2$ \& W. God\l owski$^2$ \\
1.Wy\.{z}sza Szko\l a Pedagogiczna, ul Armi Krajowej 13/15,
42-200 Cz\c{e}stochowa, Poland.\\
2.Obserwatorium Astronomiczne Uniwersytetu Jagiello\'{n}skiego, ul. Orla 171,
30-244 Krak\'{o}w, Poland \\
(E-mail:bogdan{@}wsp.czest.pl \& godlows{@}oa.uj.edu.pl)}
 
\date{}
 
\maketitle
 
\begin{abstract}
   We divide  some of the observed  diffuse
interstellar bands (DIBs) into families which appear  to have
 spectral structures of single species. Three different methods are
 applied to separate such families, exploring the  best approach  for
future investigations of this type. Starting with statistical treatment
of the data, we found that statistical methods by themselves give
insufficient results. Two other methods of data analysis
("Averaging EWs" and "Investigating the figures with arranged spectrograms")
were  found to be more useful as tools for finding the spectroscopic families
of DIBs. On the basis of these methods, we suggest some candidates
as "relatives" of 5780 and 5797 bands.
\end{abstract}
 

\section{Introduction}
 
   Diffuse interstellar bands (see e.g. Herbig  1995),  absorption
structures  of  interstellar  origin   still   await
explanation. The identification of the carrier(s) of  DIBs is
one of the most difficult challenges for spectroscopists.
 
   To the present day, huge amounts of  observational  data  on
DIBs have been stored  by  astronomers  and  published  in  hundreds  of
papers.  Unfortunately,  astronomical  data  still  do  not   meet
sufficient   understanding   from  the point of view of laboratory
spectroscopists.
 
   One tries, in general, to solve the mystery of  the  carrier(s)
of  DIBs  on  the   field   of   interdisciplinary   spectroscopic
collaboration between molecular physicists, molecular chemists and
astronomers. One expects that some progress on  this  field  will  be
possible when all known DIBs (about  three  hundred  of  them have been
discovered in visible light region) are divided  into  families  in
such a way that only one carrier  is  responsible  for  all  bands
belonging  to a given  family.  Such  families  of  bands  we  call
'spectroscopic' ones, to distinguish them from 'characterological'
families isolated by the other authors (Chlewicki et.al. 1986,
Kre{\l}owski \& Walker 1987, Josafatson \& Snow 1987). All bands belonging
to the spectroscopic family are, by  definition,  caused  by  the  same
carrier. Bands belonging to the characterological family have some
common characteristics (e.g. all are rather narrow),
but  they  may  be  produced  by  different carriers.
 
   To isolate spectroscopic families of bands, first of all one has
to by-pass in some way the problem of the "noisy correlation".
  In  a  given  spectrogram  we  have  to  deal  with
a complicated mixture of interstellar absorption structures. This is
because the medium between the target star and the  observer  contains
various species. For different directions (various  target  stars)
we have to deal with different column  densities  of  interstellar
matter giving contributions to the  spectra.  Intensities  of  all
spectral lines (bands) measured in a spectrogram are well correlated
with  the  column  density  of  relevant  matter and therefore also lines
originated by different species are mutually correlated. Such correlation
we call "noisy correlation". Of course not only differences in column
densities of relevant interstellar matter may produce noisy correlation
between DIBs. Other contribution to correlation of this kind may be given,
for example, by mutually dependent astrochemical processes.
Noisy correlation tells us nothing about spectroscopic families of
bands.
 
   The number of recognized DIBs has  grown  dramatically  in  recent  years,
primarily due to better quality of  observational  material.  More
and more weak DIBs (WDIBs)  seem  to  appear  every  time a  given
spectral region is analysed carefully. One of the authors of  this
paper (BW) spent a few years analysing spectra  for  new,
very weak DIBs. It turned out to  be  of  great  importance  to
reexamine the problem of DIB families. The aim of this paper is to
explore further the properties of DIBs in the context of isolating
families  of  the  structures.  We  first  describe  observational
material which  revealed  the above mentioned  absorption  bands.  Then  we
present the results  of the measuring  procedure  and  describe  DIB
searching methods. In the last section of the paper we discuss the
problem of some adequate method for separating spectroscopic  families
of DIBs, and we pick out two probable "relatives" for 5780 band and four other
ones for 5797 band.
 
\section{Observational material}
 
\begin{table*}
\vskip 22cm
\includegraphics{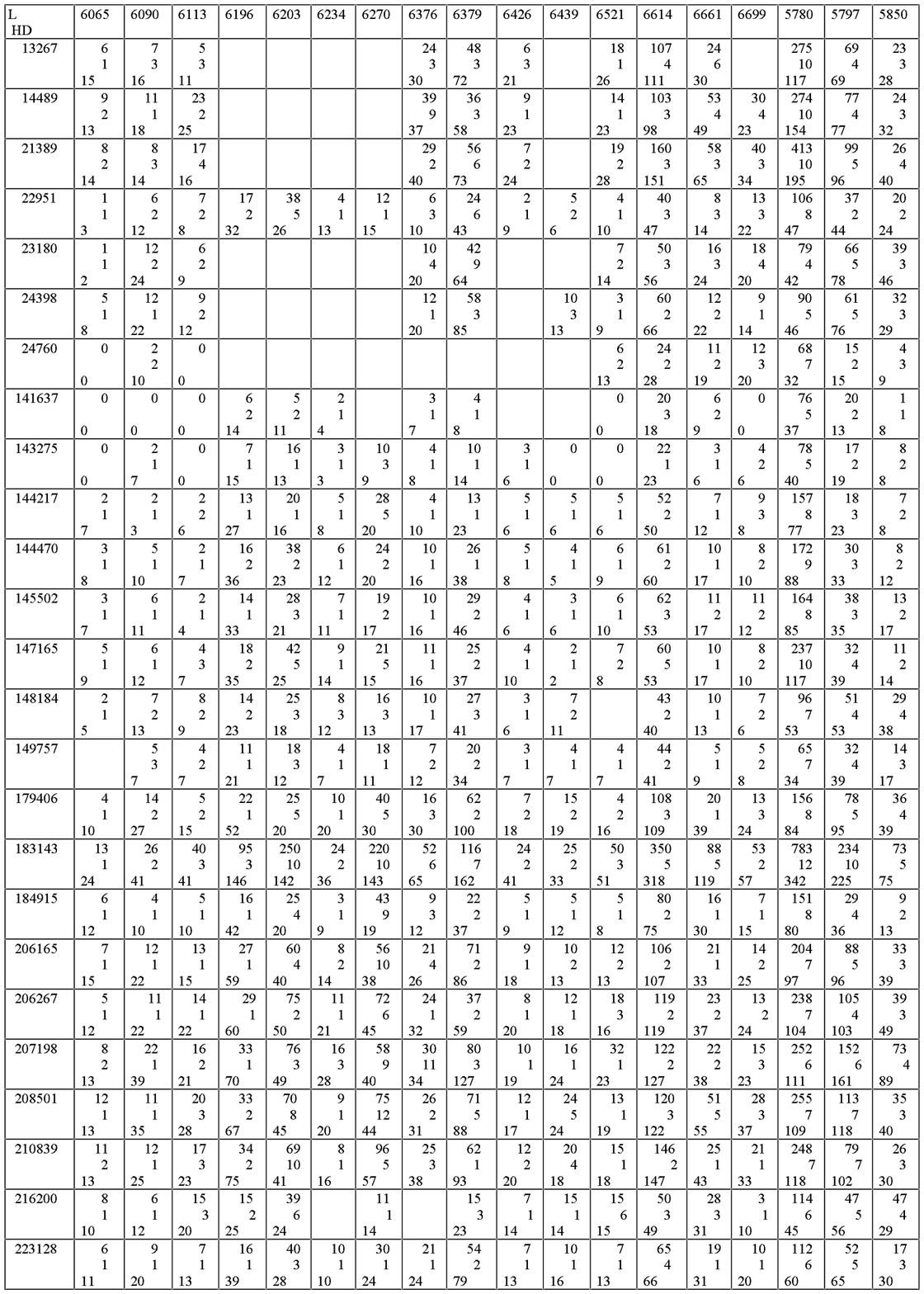}
\caption{Measured strengths of DIBs in the McDonald data. HD  numbers
for target stars are given in the first column. The first row of the table
contains  the  names  of  DIBs  (their   approximate   positions).
Individual  cells  contain  subsequently  (going  from  the  top):
equivalent width (EW), roughly estimated measuring  error  for  EW
(EW and its error are given  in  milliangstroms)  and  line  depth,
given in promilles of the local continuum level.  Empty  cells  in
the  table  indicate  to  the  cases  of  very  bad   quality
spectrograms.}
\end{table*}
 
  All the spectra analysed for the  purpose  of  this  paper  were
taken  from  the  archives  of Prof. Jacek  Kre\l owski   (Astronomical
Center, Nicolaus Copernicus University, Toru{\'n}, Poland). We used
spectra taken with the Canada-France-Hawaii Telescope (CFHT) (described
by Snow and Seab (1991)) covering  the  spectral  range  5780-5905
{\AA} and spectra acquired at the  McDonald  Observatory  with an
echelle spectrograph fed with the 2.1 m  telescope,  covering  the
spectral range  from  5600  to  7000 {\AA}  (Kre\l owski  and
Sneden, 1993). Spectra acquired with the CFHT have  S/N  of  about
800 and resolution R of about 60000. The relatively  high  quality  of
the observational data used allowed us to measure the  equivalent
widths for many strong and weak DIBs.
 
\section{ Measurements}
 
   Firstly, from the McDonald  Observatory  data,
spectrograms of the highest possible quality were selected for 25
targets. Using the
chosen spectra, we  measured equivalent widths (EW) end  depths
of 18 prominent DIBs within the spectral range from 5760  to  6700
{\AA}. HD numbers of  stars  and  designations  of  considered
DIBs, as well as values of measured parameters are given in  Table 1.
In each cell of the table (provided it is not empty) we have (from
top  to bottom): EW (in milliangstroms), error of  the  measurement  of
the EW, and line depth (in promilles of continuum). In  the  cases
of problematic measurements,  due  to  unsatisfactory  quality  of
spectra, the cell in the table was left empty.
 
\begin{table*}
 
\noindent
\begin{tabular}{crr}
WDIB's Name&Position [\AA]&Position [$cm^-1$]\\
 $5760$&$ 5760.39$&$ 17359.94$\\
 $5763$&$ 5762.72$&$ 17352.92$\\
 $5766$&$ 5766.08$&$ 17342.80$\\
 $5769$&$ 5769.11$&$ 17333.70$\\
 $5773$&$ 5772.58$&$ 17323.28$\\
 $5776$&$ 5775.87$&$ 17313.41$\\
 $5793$&$ 5793.19$&$ 17261.65$\\
 $5795$&$ 5795.11$&$ 17255.93$\\
 $5809$&$ 5809.27$&$ 17213.87$\\
 $5819$&$ 5818.75$&$ 17185.82$\\
 $5829$&$ 5828.52$&$ 17157.01$\\
\end{tabular}
 
\caption{The list of names and exact positions for WDIBs. Positions
were measured in averaged spectra (Fig.1)  and  their  errors,  in
most cases, should be less than 0.1 \AA.}
 
\end{table*}
 
\begin{table*}
\vskip 22cm
\includegraphics{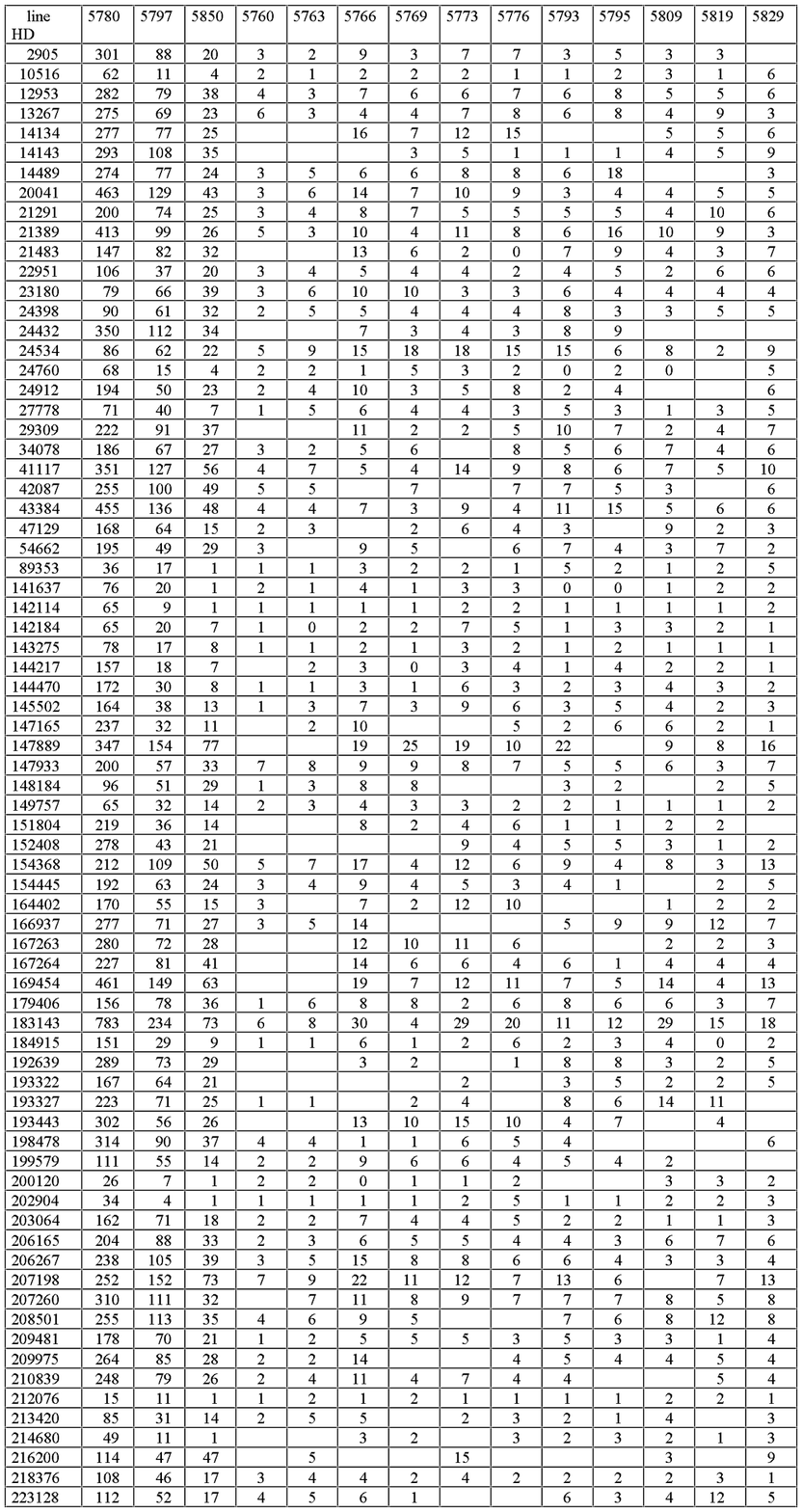}
\caption{Measured EWs (in m\AA ) for WDIBs listed
in  Table  2  and  for  three
prominent bands - 5780, 5797 and 5850. Empty cells correspond to
cases of quality of spectrograms too bad to find  any  weak  band.
Zero values were put in for non-remarkable bands in spectrograms of
quite good quality.}
\end{table*}

   We have also selected spectra  (McDonald)  covering  spectral
range from 5750 to  5860 {\AA}  for  74  stars  and  we  have
measured EWs for 11 WDIBs (listed in Table  2)  as  well  as  for
three well known and relatively strong DIBs, namely 5780, 5797 and
5849. Results of the measurements are given in Table 3. Empty cells in
the table correspond to situations where the measuring procedure was
impossible to carry out. Errors for EWs in the case of WDIBs are of
course large and in many cases they correspond to the measurement value.
 
   Positions  of  WDIBs  given  in  Table  2  were   measured in a
spectrogram which was averaged over 48 stars (in the case of  5760
and 5763 averaging was over 3 stars) observed  with  CFHT.  Before
performing the averaging procedure, all  points in individual spectra
were shifted along the $\lambda$-axis  a value
$\Delta \lambda$ ($\Delta \lambda /\lambda =const$) calculated in
such a way that the position of the interstellar NaI  D2  line  became
the same as  in  laboratory  ($\lambda =5889.95$).  All  measurements
were carried out by one of the authors (BW) and by  using  the  same
measuring procedure in all cases. The averaged spectrum compared with
the single spectrum of star HD147165  is shown on Fig.1.
 
\begin{figure*}
\vskip 22cm
\includegraphics{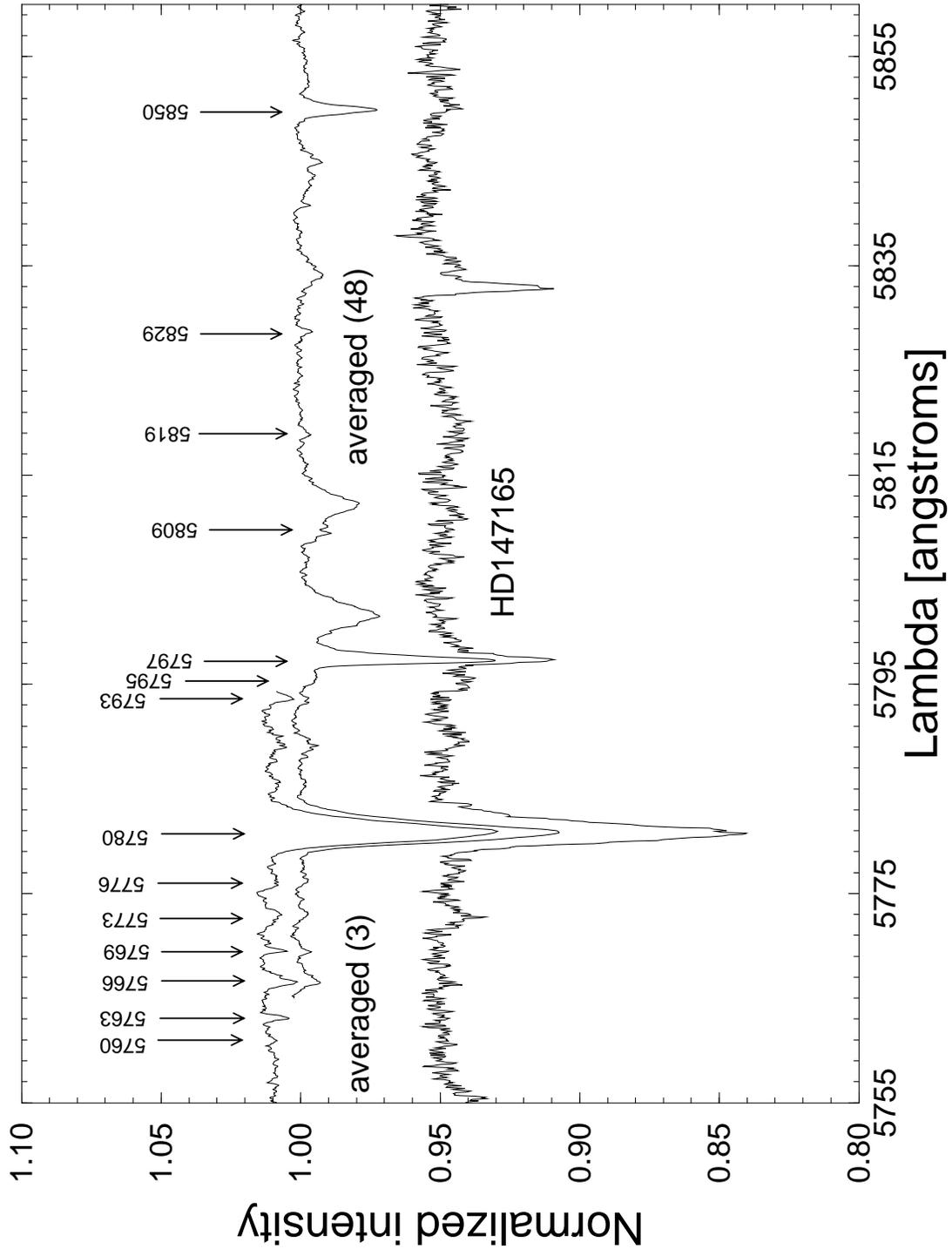}
\caption{Averaged (over different stars) CFHT spectrograms. Vertical
arrows indicate positions of studied bands. The  shorter  piece  of
spectrogram  is  a  result  of  averaging  over   3   stars   (but these
spectrograms were of the highest signal  to  noisy  (S/N)  ratios)
whereas the longer one gives an average over  48  spectrograms  of
moderate S/N ratios. Spectrograms used in the averaging procedure look
like this one for HD147165, given here for comparison.}
\end{figure*}

\section{ Searching the WDIBs}
 
   The reality of the  considered  WDIBs  and  their  interstellar
origin were checked carefully long ago by one of the authors (BW)
 and  then it was  also
claimed by other authors (e.g. Kre\l owski  and  Sneden 1993,
Galazutdinow et.al. 2000).
The routine tests usually used to prove the interstellar origin  of
the absorption lines were described in the quoted papers in detail. Here we
focus our attention  only  on  positions  and  rough profile
characteristics of  WDIBs considered.
 
   The knowledge of WDIB profiles is of  fundamental  importance;
firstly, because we need their shapes to measure EWs more precisely,
and secondly, to make spectroscopic investigations easier  (mainly
the comparisons with laboratory  spectra  of  different  molecular
samples). To study the WDIB profiles one needs spectrograms of
very  high  resolutions  as  well  as high  S/N   ratios. Unfortunately,
our data are not as good as required for such a purpose. To gain
higher S/N ratio in the spectral range occupied by considered WDIBs we
have averaged CFHT spectra over different stars. Averaged  spectra
obtained are displayed on Figure 1. Using these  spectra  we  have
measured positions of the WDIBs (Table 2). Measurements of the EWs
for  WDIBs  were  carried   out   keeping   in   mind their profile
characteristics, visible in Figure 1. Profile characteristics visible
in averaged spectrum were helpfull during EW measuring procedure for
WDIBs in individual spectra with unresolved left and right limits
of the profile.

\section{ Statistical analysis for strong DIBs}

   As far as relatively strong bands are concerned we  have
analysed  statistically mutual  dependence  between
EWs, and separately between depths, for  18  prominent DIBs (data from Table 1).
 Because  all  considered
bands are very far from  being  saturated,  the  members  of  one
spectroscopic family should  be linearly correlated.
 
   One of the most important tests giving information about linear
dependence between two variables is the test for the existence  of
linear correlation between these variables.  Therefore, we  obtained
coefficients of linear correlation $r$, for all considered pairs of bands.

\begin{table*}
\vskip 12cm
\includegraphics{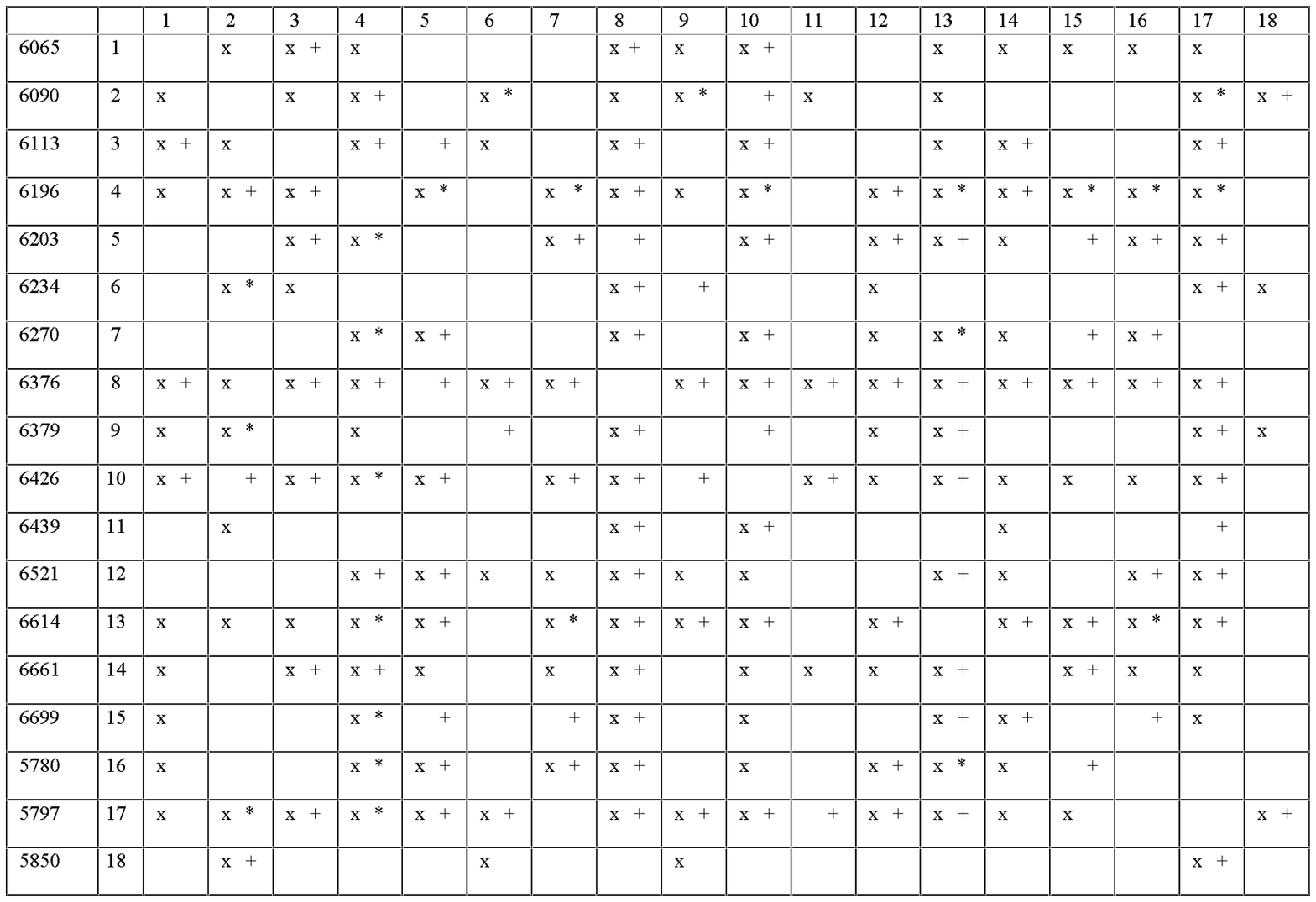}
\caption{Line-line correlations between  prominent  DIBs  listed  in
Table 1. The sign "+" means that EWs of considered lines are  well
correlated, with the value of coefficient of  linear  correlation,
$r$, greater than 0.77. When this coefficient is  greater  than  0.9
the sign "*" is used. The sign "X" stands  for  good
correlation ($r>0.79$) between the depths of bands.}
\end{table*}

Results of the study of EWs and depths for  18  prominent DIBs are given
in  Table 4.
Preliminary analysis  show that all prominent DIBs are mutually
dependent. Thus, during our search for possible spectroscopic families,
we should concentrate on the cases of extremely high correlations.
Marks in  Table 4 were put only
for these cases of correlation for which the value of statistic $t$
(see Brandt 1970):
      $$
t={R \over \sqrt {1-R^2}} \sqrt {n-2} \qquad , \nonumber
      $$
(where $R$ is a value of the parameter of the linear correlation obtained
during estimation)  distributed according to Student  distribution  with
$n-2$ degrees  of freedom,
was twice as much as that obtained for the confidence level 0.995.
 
Theoretically on the basis of results contained in Table 4 we would be able
to select some
DIB families. However, by inspection of Table 4, we find for example, that
the 6196 band has in its family also the following bands: 6203,  6270,  6426,
6614, 6699, 5780, 5797. But the 5780  and  5797 bands do not belong
to the same spectroscopic family as was shown by Kre\l owski and Westerlund
(1988). This indicates clearly that DIBs could be quite well
 correlated even if they do not belong to the same spectroscopic family,
and that we cannot rely only on a formal statistical approach when looking for
such families of DIBs. It is very probable that abundances of different agents
of DIBs are relatively well correlated only due to
properties of interstellar matter. Physical conditions which allow one
molecule to be formed within interstellar clouds simultaneously may give
a chance for the other ones to also originate. In such  cases certain noisy
correlation has to appear.

\section{ Looking for correlations between WDIBs and 5780 and 5797}
 
  The extremely low EWs of WDIBs (in most cases less than 10 m\AA)  make
their measurements very difficult. Avoiding  large  errors  in the
measured parameters is impossible. In such a case  one  looks  for an
adequate  method  to  find  any  possible   correlations   between
individual WDIBs and  chosen  strong  DIBs. As it was shown in section 5,
statistical correlations do not always give us  entirely  satisfying  results
even if the considered bands are relatively strong. As already
mentioned,  5780 and 5797  bands  do not belong to the
same spectroscopic family (Kre\l owski and Westerlund 1988).
Each of these strong bands may therefore play a role  of  a
representative band for its own family and may have its "relatives"
among WDIBs.
 
\subsection{ Statistical  analysis  of  the  intercorrelations
 between WDIBs and prominent DIBs}
 
   We  have  analysed  statistically  mutual  dependence  between
strong and weak DIBs (data from Table 3). In some cases we obtained from
measurements value zero or had a lack of measurements for WDIBs. We excluded
all such cases from the analysis. Computed  coefficients  of   linear
correlation between EWs of each of the three  strong  bands (5780,
5797, 5850) and EWs of WDIBs are presented in Table 5.
In all cases (with only  one exception 5780 versus
5769) the parameter of linear correlation $r$ is greater than zero
at a high significance level (for level of confidence 0.95 the critical value
of $r=0.23$).
 This means that EWs of each of the strong bands are mutually dependent
(in the statistical sense) and correlated.
 However, it  should  be  noted  that  in  some cases, even if
 the value of the parameter $r$ is  high enough  according to our criteria,
 the  $\chi^2$  test suggests that a more  complex  dependence,
rather than the linear one, may be relevant.
From the detailed analysis  we obtained:
(i) WDIBs 5760, 5763, 5809, 5819 are linearly-dependent with all considered
strong bands, as was checked using  the $\chi^2$ test,
(ii) in the case of strong bands 5797 and 5850
we can also accept the hypothesis of the linear regression fit for the
5793 and 5829 WDIBs, contrary to the strong band 5780 for which such
hypothesis does not work,
(iii) for the other bands (i.e. 5766, 5769, 5773, 5776 and 5795)
the hypothesis of linear dependence with all considered strong
bands should be rejected.

\begin{table*}
 
\noindent
\begin{tabular}{crrr}
&$5780$&$5797$&$5850$\\
$5760$& 0.36 &  0.54 &  0.52 \\
$5763$& 0.24 &  0.55 &  0.61 \\
$5766$& 0.67 &  0.79 &  0.74 \\
$5769$& 0.22 &  0.48 &  0.58 \\
$5773$& 0.69 &  0.70 &  0.66 \\
$5776$& 0.67 &  0.59 &  0.50 \\
$5793$& 0.49 &  0.76 &  0.81 \\
$5795$& 0.63 &  0.53 &  0.40 \\
$5809$& 0.68 &  0.68 &  0.55 \\
$5819$& 0.56 &  0.61 &  0.49 \\
$5829$& 0.51 &  0.77 &  0.84 \\
 
\end{tabular}
 
\caption{The correlation table between strong DIBs  (listed  in  the
first row of the table) and WDIBs (listed in the first  column  of
the  table).  Formal  coefficients  of  linear  correlation   were
calculated on the basis of the data taken from  Table 3.}
 
\end{table*}

The result in point (i) clearly shows that some of the WDIBs are
linearly well correlated with different DIBs (5780, 5797) which
definitively do not belong to the same spectroscopic family. Therefore, the
 statistical approach to our data is not sufficiently strong by itself
 to discriminate the membership of WDIBs into families. More
detailed analysis, taking into account other methods, is therefore required.

\subsection{ Averaging EWs method}
 
   To check mutual dependence between  WDIBs  and  5780  and  5797
bands, we have started from the following procedure. From Table 3
we extracted a sample with EW(5797) restricted  to a very  narrow
range  of  values.  From  such  a  sample  we  separated  two
subsamples: the first one with relatively low, and the second  one
with relatively high values of EW(5780). In  these  subsamples  we
calculated mean values of EW(5780), EW(5797) and  EWs  for  WDIBs.
Both subsamples, as well as results of averaging, are displayed in
Table 6. WDIBs correlated with 5780 should follow  changes  in
EW(5780), contrary to those which are not correlated.

\begin{table*}
\vskip 9.5cm 
\includegraphics{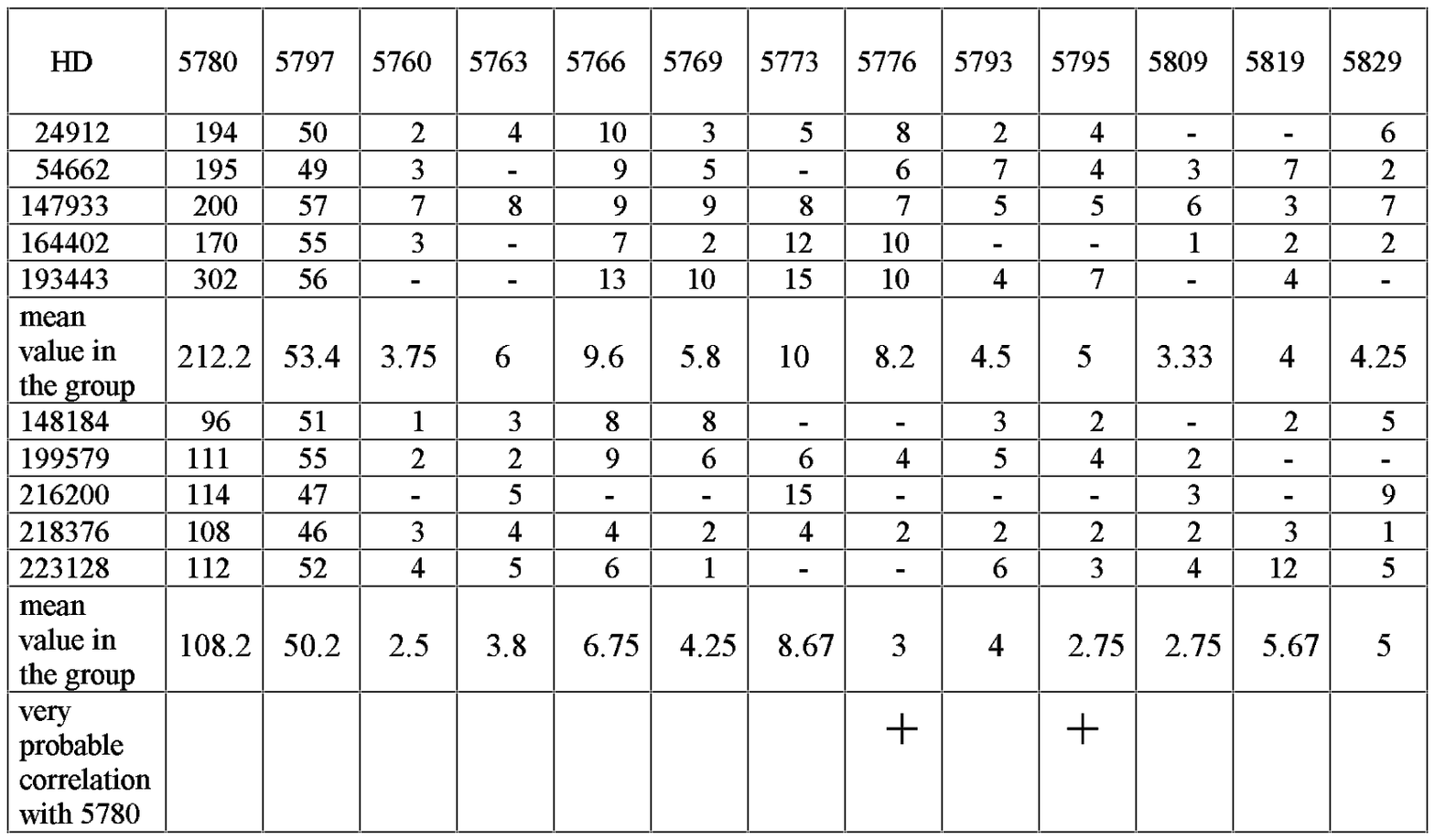}
\caption{Two subsamples of stars chosen in such a way that  average
EW(5797)  is  almost  constant  in  both  subsamples,  contrary  to
EW(5780) which in the first subsample is almost twice as large as  in
the second one. The names of bands are listed in  the  first  row,
whereas names of target stars used are given in the  first  column
of the table. In both subsamples for each band the mean  value  of
measured EWs was counted. The sign "-" stands for the lack of  the
data and "+" indicates interesting examples  of  correlation  (see
text).}
\end{table*}
 
   In a quite similar way we have extracted a sample of stars with
almost  constant  EW(5780)  and  separated  subsamples   with
opposite values of EW(5797) - Table 7.  Such  a  simple  procedure
allowed us to divide WDIBs into three groups; (i) well  correlated
with 5780 (5776, 5795), (ii)  well  correlated  with  5797  (5793,
5819, 5829) and (iii) correlated with neither  5780  nor  5797
(5769, 5773, 5809). Three  other  cases  (5760,  5763,  5766)  are
rather complicated because they seem to  be  correlated  with
5797 as well as with 5780. The main selection criterion  here was a
gradient  of the average  value  for  WDIBs.  If  this  gradient, for
a considered WDIB, is evidently similar  to  the  gradient  which  is
typical for the leading band (in our case this gradient is very close
to 0.5) we  included this WDIB into the family.
 
\begin{table*}
\vskip 10cm
\includegraphics{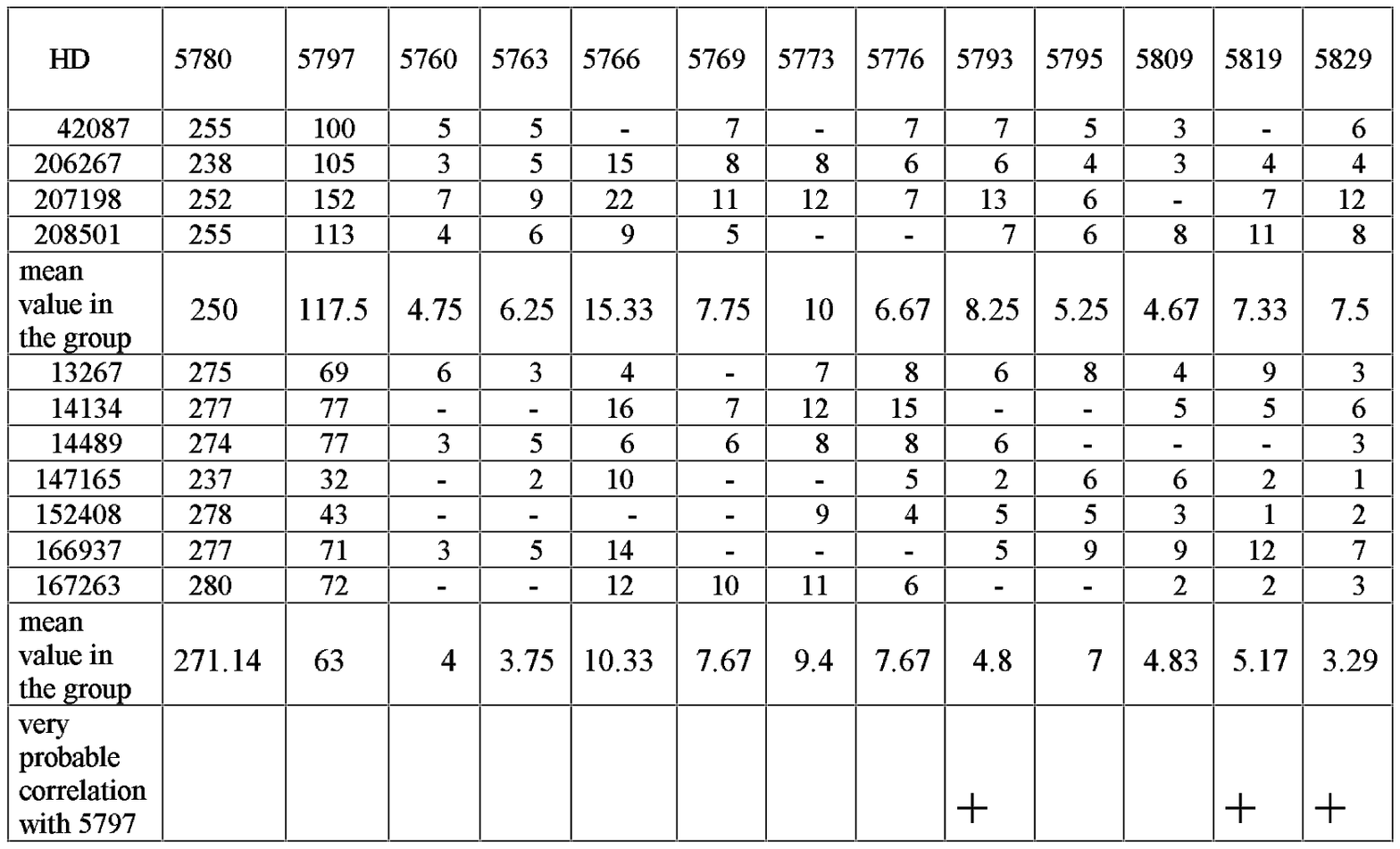}
\caption{The same as in Table 6  but  here the  mean  EW(5780)  is  rather
stable whereas EW(5797) drops almost in half  when  going  from  the
first subsample to the second one.}
\end{table*}

  Please  note that the term "correlation"  used  above  does  not
mean the same as correlation in the statistical sense.  Statistically,
all considered bands are quite well mutually correlated.
 
\subsection{ Investigating the figures with arranged spectrograms}
 
   Another way to divide  WDIBs  into the three  above  mentioned
groups consists in visual inspection of  spectrograms. If we compose
sequences of  spectrograms  in  such  a  way  that  EW(5780),  for
example, is constant and EW(5797) gradually increases  (Fig.2)
we are able, after visual  inspection  of  the  picture, to indicate
WDIBs which follow, and  those  ones  which  do  not  follow,  the
behaviour of 5797. Figure 3  shows  the  case  of  nearly  constant
EW(5797) and changing EW(5780).
 
\begin{figure*}
\vskip 22cm
\includegraphics{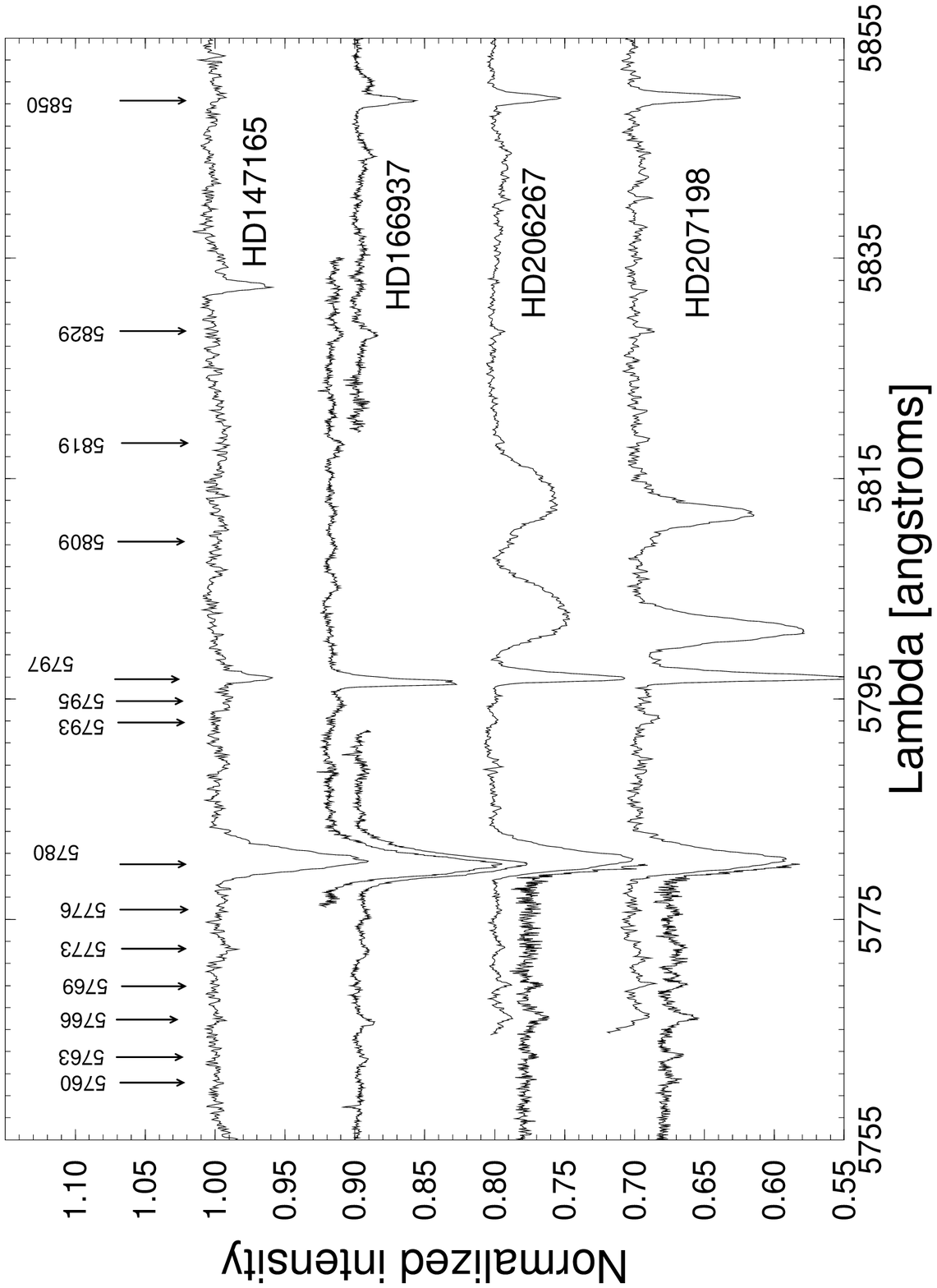}
\caption{CFHT  (the  longest  pieces)  and   McDonald's   (shorter,
individual orders of echelle spectrograph)  spectrograms  arranged
in such a way that strength  of  5797  band  gradually  increases
whereas intensity of 5780 band remains constant.}
\end{figure*}

\begin{figure*}
\vskip 22cm
\includegraphics{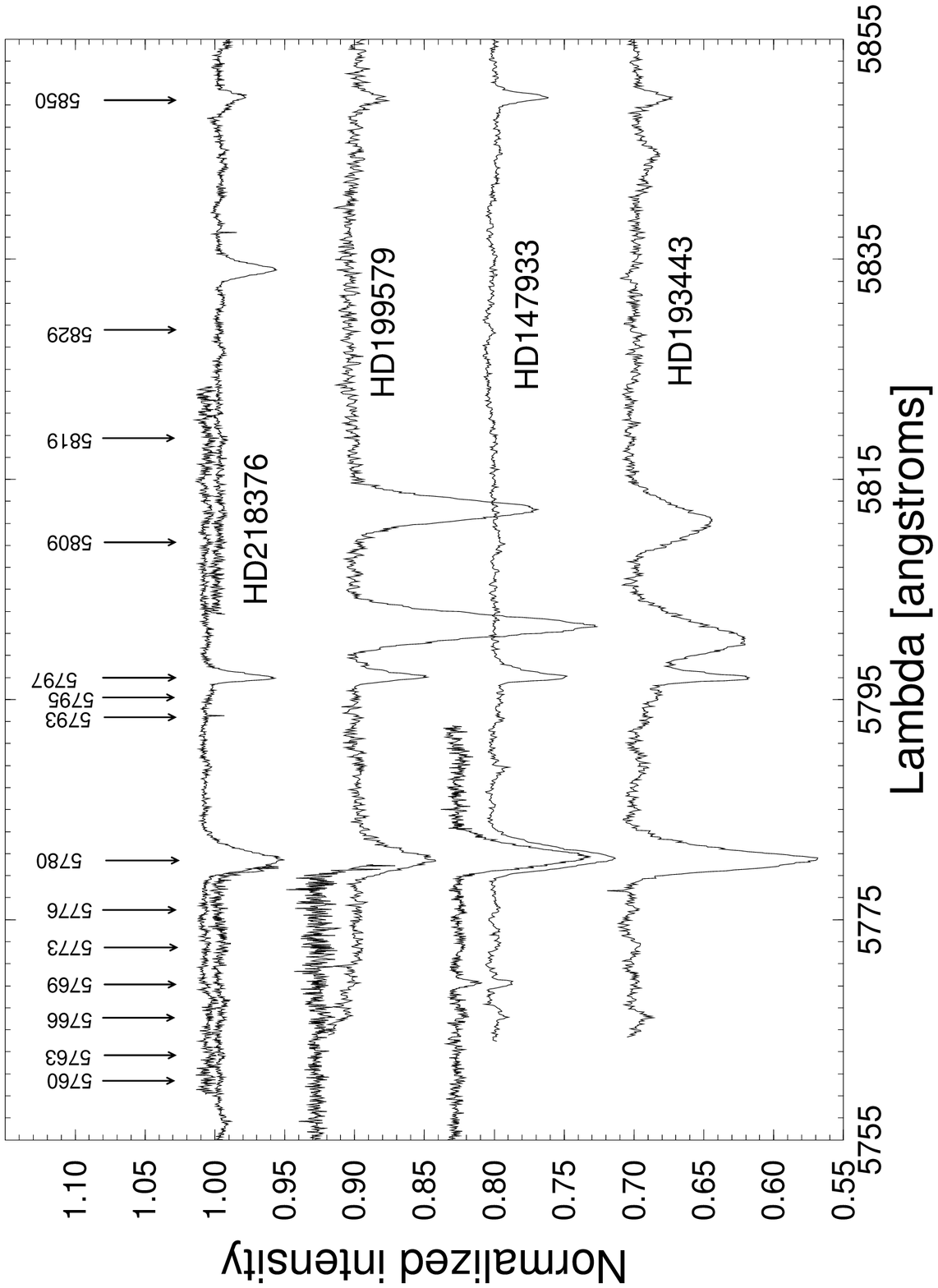}
\caption{Similar to Figure 2, but here the roles of 5780  and  5797  bands
are exchanged. Intensity of the first band increases  whereas  the
strength of the second one remains constant.}
\end{figure*}

  This inspection procedure of Figures 2  and 3
confirms  results  achieved  with the earlier   described   averaging
procedure. Furthermore, one can, when studying Figures 2 and  3,
make the following conclusions:
 
 - the 5850 band, which corresponds to relatively strong phenomena, is very
well correlated with the 5797 band and not (at all) with the 5780 one  (this
was claimed already by Kre\l owski {\it et al.}, 1993),
 
 - the 5763 band follows the behaviour  of  the  band  5797  and
does not follow (at all) the 5780 band,
 
- although Figures 2 and 3 are not as good as required for  studying
the 5760 band it seems that this WDIB is  correlated  rather  with
5797 than with the 5780 band,
 
- the 5769 band also seems to be well correlated with the 5797 band.
 
   It is worthly of note that the observations  we  are  dealing  with
here were not planned for searching for spectroscopic  families.  One
may, however, perform much more adequate observations,
directed to the subject.
In the case of better data, similar  analysis  would  give
much more reliable results and lead directly to separation of true
spectroscopic families of bands.

\section{ Discussion}
 
   As   mentioned  in the introductory  section,  the  main
obstacle to isolating spectroscopic families of DIBs is the  noisy
correlation. Due to a very high level (as the analysed  data reveals)  of
noisy correlation in the considered data, the ability of statistical
method to isolate spectroscopic families  are  very  limited.
 
   Tight linear correlation, expected between members of the  same
spectroscopic family, is efectively hidden by noisy  correlation
and by measuring errors.  This  is evident  when  we  study
results of statistical analysis  described  in the  previous  section.
Looking at Table 5 only, we are not able to indicate which WDIB
belongs to the same spectroscopic family as, for example, the  band  5797.
On the other hand, when considering Table 4, we would be inclined to
isolate, e.g, the family of strong lines: 6196, 6203,  6270,  6426,
6614, 6699, 5780, 5797; and this would be a mistake, since 5780  and  5797
belong to different families, as  mentioned in section 5.
 
   Taking into account that statistical analysis requires  plenty
of usable data and gives insufficient (for solving our problem by itself)
results, it  is therefore not recommended as an appropriate tool for isolating
spectroscopic families. [However, the statistical approach may be useful
to distinguish linearly correlated bands  from these ones  which  are
correlated  non-linearly.  Non-linearly  correlated  DIBs   should
belong to different spectroscopic families. Also multidimensional
statistical analysis could be useful in this case.
Further study of this problem will be the subject of a separate paper
(God\l owski \& Wszo\l ek, in preparation)].
 
   Methods described in subsections 6.2  and  6.3  are  much  more
appropriate than the formal statistical aproach. They need a fewer  number
of spectrograms and much  less  time  for  making  EW  (or  depth)
measurements. These methods seem  also  to  give  valuable
results. Using these methods we performed preliminary separation of two
presumable spectroscopic families: (i) 5780, 5776  and  5795,  and
(ii) 5793, 5797, 5819, 5829 and 5850.
 
   Most probably, designated families are not complete yet. One cannot
exclude also the possibility that we made wrong indications.
In the case of almost constant ratios between column densities  of
various DIB carriers in interstellar clouds, we have a chance to  get
results quite similar those of the case when we have to deal with few
spectral lines of the same  carrier. Further  investigation,
based on better data samples and  involving  other  spectral
ranges, is necessary to isolate  true  spectroscopic  families  of
bands.
 
\section*{ Acknowledgement}
  We are grateful to Prof. David Batuski and Dr Krzysztof Maslanka
who have corrected grammar in the manuscript.
BW would like to thank Prof.  Jacek  Kre\l owski
for giving him access to his data archives, as well as to his
specialised software.


\begin{thebibliography}{}
\bibitem[\protect\citename{aa}1993]{a}
Brandt S., 1970, Statistical and Computational Methods in Data Analysis.
Nord Holland Publishing Company
\bibitem[\protect\citename{aa}1993]{a}
Chlewicki G.,van der Zwet G.P.,van Ijzendoorn L.J.,Greenberg J.M.,
1986, ApJ. 305,455
\bibitem[\protect\citename{aa}1993]{a}
Galazutdinow G.A.,Musaev F.A.,Kre\l owski J., Walker G.A.H., 2000, PASP,112,648
\bibitem[\protect\citename{aa}1993]{a}
Herbig G.H.,1995, Ann.Rev.Astron.Astrophys. 33,19
\bibitem[\protect\citename{aa}1993]{a}
Josafatson K.,Snow T.P.,1987, ApJ. 319,436
\bibitem[\protect\citename{aa}1993]{a}
Kre\l owski J.,Sneden C.,1993, PASP 105,1141
\bibitem[\protect\citename{aa}1993]{a}
Kre\l owski  J.,Snow  T.P.,Papaj   J.,Seab   C.G.,Wszo\l ek   B.,1993,
ApJ. 419,692
\bibitem[\protect\citename{aa}1993]{a}
Kre\l owski J.,Walker G.A.H.,1987, ApJ. 312,860
\bibitem[\protect\citename{aa}1993]{a}
Kre\l owski J.,Westerlund B.E.,1988, A{\&}A. 190, 339
\bibitem[\protect\citename{aa}1993]{a}
Snow T.P.,Seab C.G.,1991, ApJ. 382,189
\end{thebibliography}
\end{document}